\documentclass[12pt]{iopart}

\usepackage{iopams}  
\usepackage{graphicx}
\usepackage{color}
\usepackage{cancel}

\expandafter\let\csname equation*\endcsname\relax
\expandafter\let\csname endequation*\endcsname\relax
\usepackage{amsmath}

\newcommand{\ignore}[1]{}

\begin{document}
\title{Tree decompositions of real-world networks from simulated annealing}

\author{Konstantin Klemm}

\address{IFISC (CSIC-UIB), Palma de Mallorca, Spain}
\ead{klemm@ifisc.uib-csic.es}

\begin{abstract}
Decompositions of networks are useful not only for structural exploration. They also have implications and use in analysis and computational solution of processes (such as the Ising model, percolation, SIR model) running on a given network. Tree and branch decompositions considered here directly represent network structure as trees for recursive computation of network properties. Unlike coarse-graining approximations in terms of community structure or metapopulations, tree decompositions of sufficiently small width allow for exact results on equilibrium processes. Here we use simulated annealing to find tree decompositions of narrow width for a set of medium-size empirical networks. Rather than optimizing tree decompositions directly, we employ a search space constituted by so-called elimination orders being permutations on the network's node set. For each in a database of empirical networks with up to 1000 edges, we find a tree decomposition of low width. 
\end{abstract}

%\maketitle

%%%%%%%%%%%%%%%%%%%%%%
\section{Introduction}
%%%%%%%%%%%%%%%%%%%%%%

% complex systems and computation 
The analysis and modeling of complex systems involves complex {\em computational} tasks. 
In comparing empirical network structures with network models, for instance, one asks if
they share common macroscopic behaviour under a percolation process or spin kinetics.

% "Non-sophisticated" networks: processes on simple graphs are difficult enough
While theory of networks in complex systems has been enriched to include adaptive \cite{Gross:2008}, temporally evolving \cite{Masuda:2016,Masuda:2013}, and multi-layer \cite{Bianconi:2018,Diakonova:2016} structures, open questions remain also for processes on simple static structures. The Ising model in the macrocanonical ensemble, for instance, is fully captured by its density of states. Computing the exact density of states is not feasible in general. A subproblem of this task is to decide if there is a state at or above a given energy $\epsilon$. Even this modest subproblem is equivalent to the NP-complete {\sc maximum cut} problem on the given network \cite{Karp:1972,Garey:1979}. 

% Community structure, stochastic block models, metapopulations: coarse-graining 
Computational problems on networks are made more feasible when effective size is reduced by coarse-graining, e.g.\ \cite{Masuda:2009}. Several nodes with similar neighborhoods forming a community \cite{Fortunato:2010} or block \cite{Doreian:2019} are lumped together into one representative node. Similar coarse-graining is applied in metapopulation models for epidemic spreading \cite{Sattenspiel:1995,Riley:2007} and opinion formation \cite{FernandezGracia:2014} where the nodes of the network are geographic locations rather than individuals. Coarse-graining amounts to lossy compression \cite{Rosvall:2007}, however, so precision of results decreases with the amount of coarse-graining. 

Here we advocate the study and development of exact methods \cite{Lucet:1999,Woeginger:2003} rather than approximations and heuristics. In parallel to the development of exact methods, their applicability is to be explored by finding the networks and classes of networks that allow for efficient computation.

Tree networks allow for particularly simple recursive methods \cite{Pearl:1982} where the network itself is used as the recursion tree. For non-tree networks, the same recursion equations are applied as a tree-like heuristic \cite{Mezard:2002,KarrerPRL:2014,HamiltonPRL:2014,Moore:2020}. Recent research generalizes message passing to account for cycles to a certain extent. Radicchi and Castellano introduce corrections due to triangles only \cite{RadicchiPRE:2016}. Cantwell and Newman consider expansion in cycle length and use Monte-Carlo sampling \cite{ToralColet:2014} to deal with combinatorial explosion in the number of terms in the recursion \cite{Cantwell:2019}.

Rather different generalizations of the tree property, termed tree decompositions and branch decompositions have been known in the area of combinatorial optimization and graph algorithms \cite{Arnborg:1987,Robertson:1991,Bodlaender:2010,Bodlaender:2011}. A network has tree-width $k$ if it can be split into smaller subnetworks with overlaps of at most $k$ nodes, where this property holds again for the subnetworks; the tree representation of this split recursion is then called a tree decomposition of width $k$. See Section~\ref{sec:background} for the strict definition. Computations with running time $O(f(n))$ for a naive (e.g.\ exhaustively enumerative) algorithm on system size $n$, typically reduce to running time $O(nf(k))$ when exploiting tree-width $k$. For the density of states of the Ising model, this is a reduction from $O(2^n)$ to $O(n2^k)$, making exact computation feasible as long as there is a tree decomposition of sufficiently low width $k$.

This suggests a two-step process as a general {\em modus operandi} for solving a computational problem on a given network. (i) Find a tree decomposition of possibly low width. (ii) Solve the original problem with a method efficiently exploiting the low tree-width. Step (i) may fail, either because the given network does not have a tree decomposition of sufficiently low width or because we are unable to find it. Finding tree decompositions of optimally low width is in itself an NP-hard problem \cite{Arnborg:1987}. Thus we resort to a heuristic such as a greedy algorithm or stochastic optimization in step (i) \cite{Kjaerulff:1992,Bodlaender:2010}.

Here we demonstrate this two-step process for a collection of small and medium-size networks frequently employed as test structures. In step (i), we find tree decompositions of low width by simulated annealing \cite{Kirkpatrick:1983}. Demonstrating step (ii), we use the tree decomposition and find the exact values of the networks' maximum Ising energy.

%%%%%%%%%%%%%%%%%%%%%%%%%%%%%%%%%%%%%%%%%%%%%%%%%%%%
\section{Mathematical background, methods, and data} \label{sec:background}
%%%%%%%%%%%%%%%%%%%%%%%%%%%%%%%%%%%%%%%%%%%%%%%%%%%%

\subsection{Tree decompositions}

Let $G=(V,E)$ be a graph with node set $V$ and edge set $E$. Let $\mathcal{B}$ a family of non-empty subsets of $V$. Each such $B \in \mathcal{B}$ is called a {\em bag}, being
$B \subseteq V$. Furthermore, let $T$ be a set of edges (unordered tuples) on $\mathcal{B}$, so that $(\mathcal{B},T)$ is a tree. Now this tree with node set $\mathcal{B}$ and edge set
$T$ is called a {\em tree decomposition} of $G$ if additionally the following two conditions hold.
\begin{itemize}
\item[(i)] For each edge $\{v,w\} \in E$, there is a bag $B \in \mathcal{B}$ so that $x \in B$ and $y \in B$.
\item[(ii)] The set of bags containing $v$ induces a connected subgraph of $(\mathcal{B},T)$.
\end{itemize}
In the literature (e.g.\ \cite{Bodlaender:2010}),  an additional requirement is that each node $v \in V$ is contained in at least one bag. This makes a difference only when there are isolated nodes; a non-isolated node must anyway appear in a bag due to condition (i).

Each graph $G=(V,E)$ trivially has a tree decomposition: simply take $\mathcal{B}=\{ V \}$, a single bag containing all nodes, and $T=\emptyset$. Tree decompositions of use for recursive computation, however, need to achieve small bag size. The {\em width} $w_\infty$ of a tree decomposition $(\mathcal{B},T)$ is the size of a largest bag, reduced by 1,
\begin{equation}
w_\infty(\mathcal{B},T) = \max\{ |B| : B \in \mathcal{B}\} - 1~.
\end{equation}
The tree-width of a graph $G$ is the minimum of width over all tree decompositions of $G$. If $G$ itself is a tree, $G$ has a tree decomposition with a bag $B=\{v,w\}$ for each edge $\{v,w\} \in E$. Thus a tree has tree-width 1.

\subsection{Elimination order}

\begin{figure}
  \begin{tabular}{cc}
    (a) & (c) \\ \\ \\
    \includegraphics[height=46mm]{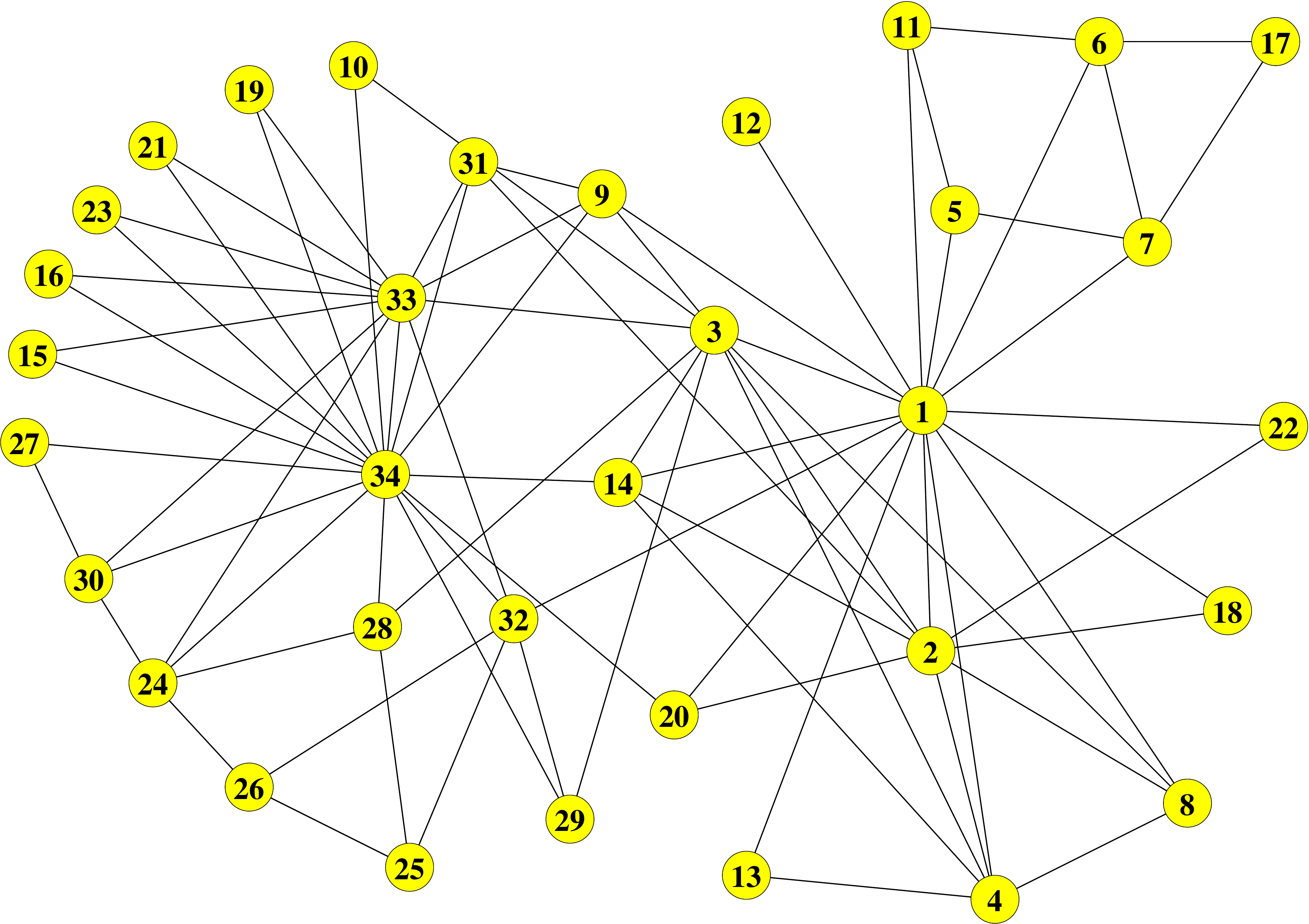} &
    \includegraphics[height=46mm]{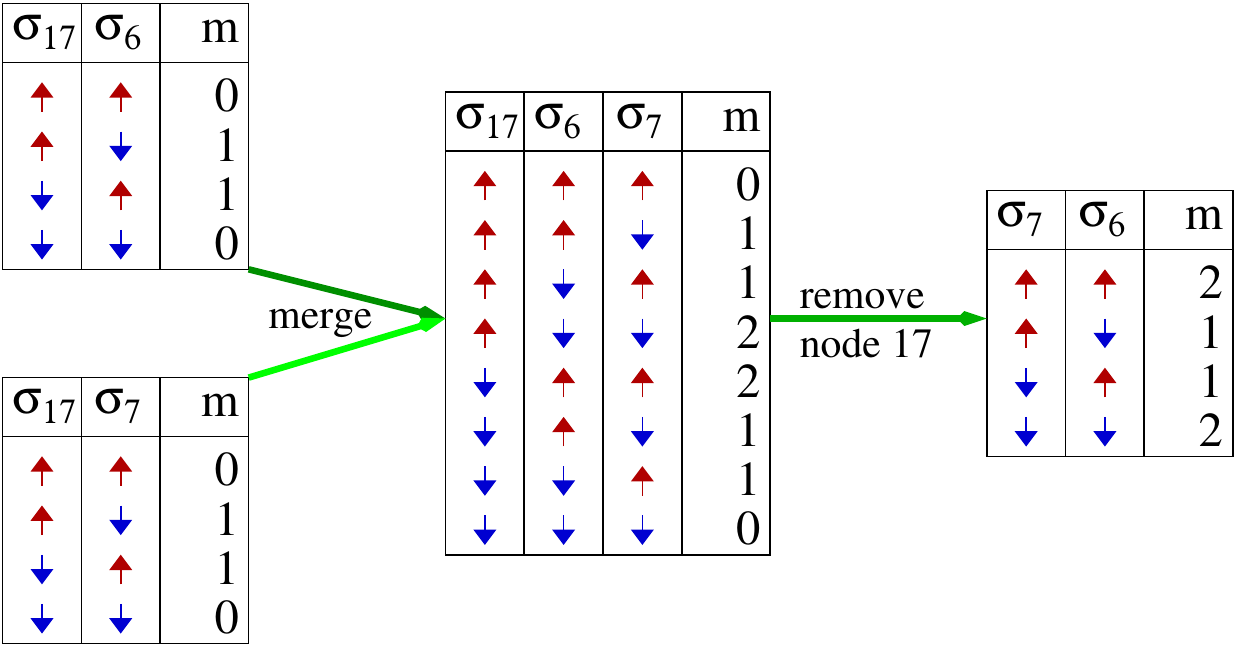}
  \end{tabular}
  \bigskip

\centerline{(b)}
\centerline{\includegraphics[height=46mm]{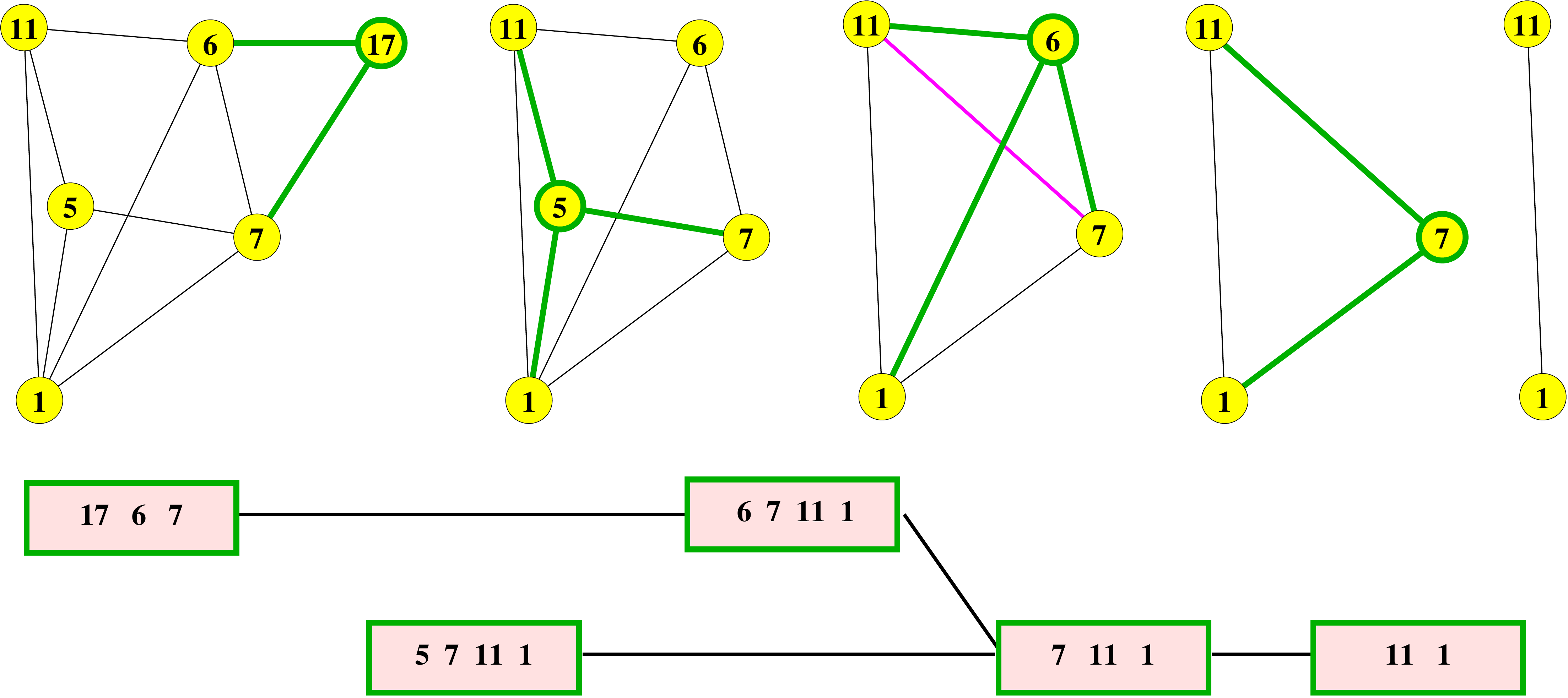}}
\caption{\label{fig:multi_illu}
(a) Karate club network \cite{Zachary:1977} to illustrate (b) elimination order and the tree decomposition generated, and (c) Initial steps of merging and removal in the computation of maximum cut size. In (b), nodes are eliminated in the order $(17, 5, 6, 7, 11, 1)$ considering only the subgraph spanned by these six nodes (upper right part of panel (a)). Thick node outline and thick edges indicate part of the network removed in each step. Note that elimination of node 5 generates and extra edge $\{7,11\}$ due to clique completion. }
\end{figure}

Finding a tree decomposition of low width for a given graph is a hard problem in itself. Stochastic search methods like annealing and genetic algorithms are useful for this task. The application of these methods is made easier by encoding the search space \cite{Klemm:12b} rather than operating on the tree decompositions themselves. Elimination order is one such encoding of tree decomposition, with the theory well explored \cite{Bodlaender:2010}.

Let us define the operation of {\em elimination} $\Lambda$. For graph $G=(V,E)$ and node $v\in V$, elimination of $v$ from $G$ generates a graph $\Lambda(G;v)$ by (i) adding all missing edges between neighbours of $v$, thus turning the neighbourhood of $v$ into a clique and (ii) removing $v$ and all its incident edges.

Now let $\pi = \pi_1,\pi_2,\dots,\pi_n$ be an ordering (permutation) of the nodes of the graph $G=(V,E)$ with $n=|V|$ nodes. For each $v \in V$, there is exactly one index $i \in n_\rfloor$ with $\pi_i=v$. The graph sequence of $G$ under elimination order $\pi$ is $G^{(0)}, G^{(1)}, G^{(2)}, \dots, G^{(n)}$ generated from initial element $G^{(0)} = G$ by recursion $G^{(i)} = \Lambda(G^{(i-1)};\pi_i)$ for $i \in \{1,2,\dots,n\}$. $G^{(n)}$ is the empty graph.

An elimination order $\pi$ for a graph $G=(V,E)$ gives rise to a tree decomposition $(\mathcal{B},T)$ of $G$ in the following way. The set of bags is $\mathcal{B}= \{B_i : i\in \{1,\dots\,n\}\}$. Now for each $i$, bag $B_i$ contains $\pi_i$ and its neighbours in $G^{(i)}$. Note that $\pi_i$ may have neighbours in $G^{(i)}$ it does not have in $G$ itself. For indices $i<j$, we have a tree edge $\{i,j\} \in T$ if and only if $j=\min\{k>i : \pi_k \in B_i\}$. This means that, for each $i\neq n$, bag $B_i$ is connected to exactly one bag with higher index, being the bag of the earliest (w.r.t.\ $\pi$) neighbour. Figure~\ref{fig:multi_illu}(b) illustrates elimination and generation of a tree decomposition from elimination order for a subgraph of the Karate club network (Figure~\ref{fig:multi_illu}(a)).

Some model networks grow incrementally by attaching a new node to an existing clique of given size $m$ \cite{DorogovtsevPRE:2001,KlemmPRE:2002a,EguiluzPRL:2002}. In this case considering the nodes in reverse order of addition yields a {\em perfect elimination order} whose application does not involve edge addition. In each elimination step, the neighborhood of the eliminated vertex is a clique already by construction. Graphs with a perfect elimination order are {\em chordal}, exhibiting further useful features \cite{Golumbic:2004}.

\subsection{Simulated annealing}

We use a standard setting for simulated annealing \cite{Kirkpatrick:1983} as a Metropolis Markov chain \cite{ToralColet:2014} with a slow decrease of temperature. We initialize the elimination order $\pi$ with a permutation drawn from the uniform distribution of permutations on the node set. In order to generate a proposal $\pi^\prime$ from $\pi$, select an index $i\in \{1,\dots,n-1\}$ uniformly at random and swap elements at positions $i$ and $i+1$:
\begin{equation}
  \pi^\prime_j =
  \begin{cases}
    \pi_{i+1} & \text{if }j=i \\
    \pi_{i}   & \text{if }j=i+1 \\
    \pi_j     & \text{otherwise}
  \end{cases}
\end{equation}

The proposal is accepted with probability
\begin{equation}
\min \left\{ \exp[ \beta(t) (r(\pi)-r(\pi^\prime))], 1 \right\} 
\end{equation}
where the cost function $r$ assigns each elimination order a real number, see section~\label{sec:costfunc}.

The inverse temperature $\beta$ depends on time $t$ as $\beta(t)=ct$ with cooling speed $c$ as parameter. Time is measured in sweeps, so the clock advances by $\Delta t = 1 / (n-1)$ in every step of proposal and acceptance.

\subsection{Cost functions} \label{subsec:costfunc} %%%%%%%%%%%%%%%%%%

A large part of work on tree decompositions focuses on minimization of tree width \cite{Bodlaender:2010}, which translates to stochastic optimization with a cost function $w_\infty(\pi)$ on elimination order $\pi$. We know of the following two problems that motivate the use of cost functions different from $w_\infty(\pi)$.
\begin{enumerate}
\item As a maximum over all bags, $w_\infty(\pi)$ only varies at moves that involve the largest bags of the tree decomposition. This renders the cost of a solution equal to that of most of its neighbouring solutions, thus lacking local gradient information in the search space.
\item While the time needed for a computational task on a network has an upper bound in terms of tree width, the time actually depends on {\em all} bag sizes $s_1, s_2, \dots s_n$ of the tree decomposition, typically as $\sum_{i=1}^n 2^{s_i}$ for spin systems.  
\end{enumerate}
Clautiaux et al. \cite{Clautiaux:2004} propose a solution to problem 1 above in terms of a modified cost function
\begin{equation} \label{eq:funcm}
m(\pi) = w_\infty(\pi) + n^{-2}\sum_{i=1}^n s_i^2~.
\end{equation}
The additional sum accounts for the sizes of all bags while the dominating term remains $w_\infty$.

We introduce the cost function
\begin{equation}
w_{\eta}(\pi)=\log_\eta [ n^{-1}\sum_{i=1}^n \eta^{s_i-1} ] ~.
\end{equation}
with a parameter $\eta>1$. This is motivated by both items 1 and 2 above. For spin systems with computation time proportional to $2^{s_i}$ for bag $i$, the parameter value $eta=2$ is to be chosen. Note that $w_{\eta}$ has the tree-width $w_\infty$ as a limit $\eta \rightarrow \infty$ if there is a unique maximum bag size.

\subsection{Solving maximum cut using an elimination order} %%%%%%%%%%%%%%

Given a network $G=(V,E)$, the maximum cut problem asks to find the largest bipartite subgraph of $G$. In other words, we are looking for a partition of the node set $V$ into disjoint subsets $V_1$ and $V_2$ to maximize the number of edges running between $V_1$ and $V_2$. Assigning set membership in $V_1$ or $V_2$ is equivalent to choosing a spin value $\sigma_v \in \{-1/2,+1/2\}$ for each node $v \in V$. Then a solution of the max-cut problem is a spin configuration $(\sigma_v)_{v\in V}$ that maximizes the number of edges between nodes with unequal spins,
\begin{equation} \label{eq:maxcut_as_spins}
H(\sigma) = \sum_{\{v,w\} \in E} | \sigma_v - \sigma_w |~.
\end{equation}
This amounts to finding the ground state of the spin glass with Hamiltonian $H$ where each edge is an antiferromagnetic bond of unit weight. Equivalently we look for a maximum energy state of the Ising (all ferromagnetic) model on the same network. The maximum cut problem is NP-hard \cite{Garey:1979}. Therefore we do not know a general solver working in time polynomial in the size of the network.

In the following, we describe an algorithm to solve the maximum cut problem using an elimination order $\pi$. 
During each phase of the computation, $\mathcal{F}$ contains a set of functions storing partial results on cut sizes. Each function $f \in \mathcal{F}$ is based on a subset $V_f$ of the node set. The domain of the function is the set $\{-1/2,+1/2\}^{V_f}$ of all spin configurations on $V_f$. Each such spin configuration is assigned the maximum cut size possible.

Initially, $\mathcal{F}$ contains one function $f$ for each edge $\{v,w\} \in V$ of the network. There are $2\times2=4$ spin configurations on the nodes $v$ and $w$. If $\sigma_v \neq \sigma_w$, then $f(\sigma_v,\sigma_w) = 1$; otherwise $f(\sigma_v,\sigma_w) = 0$.

Now the computation proceeds in a loop, running loop index $i$ from 1 to $n=|V|$ and using the given elimination order $\pi=(pi_1,pi_2,\dots, \pi_n)$. In each round of the loop
\begin{enumerate}
\item Find all functions $f \in \mathcal{F}$ with $\pi_i \in V_f$ and replace them by the merged function $f^\prime$. See below for details of merging.
\item Remove node $\pi_i$ from the function $f^\prime$ generated at step 1. See
below for details of node removal. 
\end{enumerate}
Note that merging in step 1 and removal in step 2 do not commute. Before removing a node $r$, all functions containing $r$ need to be merged into a single one.

\paragraph{Merging functions.} Two functions $f,g \in {\mathcal F}$ are merged into function $h$ having the joint node set $V_h = V_f \cup V_g$. 
For each spin configuration $\sigma=(\sigma_v)_{v\in V_h}$,
\begin{equation}
h(\sigma) = f(\sigma)+g(\sigma)~.
\end{equation}
This describes the merging of two functions. For merging more than two functions, as happens in general at step 1 above, merging is applied several times. The operation is commutative, i.e.\ the result does not depend on the order of merging.

\paragraph{Removing nodes from functions.} For a function $f \in {\mathcal F}$ and node $r \in V_f$, the removal of $r$ from $f$ generates a function $f^{\downarrow r }$ with $V_{f^{\downarrow r}} = V_f \setminus \{r\}$. For each spin configuration $\sigma=(\sigma_v)_{v \in V_{f^{\downarrow r}}}$,  we have 
\begin{equation}
f^{\downarrow r}(\sigma) = \max\{f(\sigma,\sigma_r=-1/2),f(\sigma,\sigma_r=+1/2)\}~.
\end{equation}
Independence from the variable $\sigma_r$ is thus obtained by taking the maximum of the edge count with respect to the two spin orientations at node $r$. Figure~\ref{fig:multi_illu}(c) illustrates the steps of merging and removal.

\subsection{Network data} %%%%%%%%%%%%%%%%%%%%%%%%%%%%%%%%%%%%%%%%%%%%

We use the set of networks compiled by Radicchi and Castellano
\cite{RadicchiPRE:2016} restricting it to those 15 with less than 1000 edges.
See Table~\ref{tab:mainres}.

%%%%%%%%%%%%%%%%%%%%%%%%%%%%%%%%%%%%%%%%%%%%%%%%%%%%%%%%%%%%%%%%%%%%%%
\section{Results}
%%%%%%%%%%%%%%%%%%%%%%%%%%%%%%%%%%%%%%%%%%%%%%%%%%%%%%%%%%%%%%%%%%%%%%

%\subsection{Comparing cost functions} %%%%%%%%%%%%%%%%%%%%%%%%%%%%%%%%

\begin{figure}
\centerline{\includegraphics[width=\textwidth]{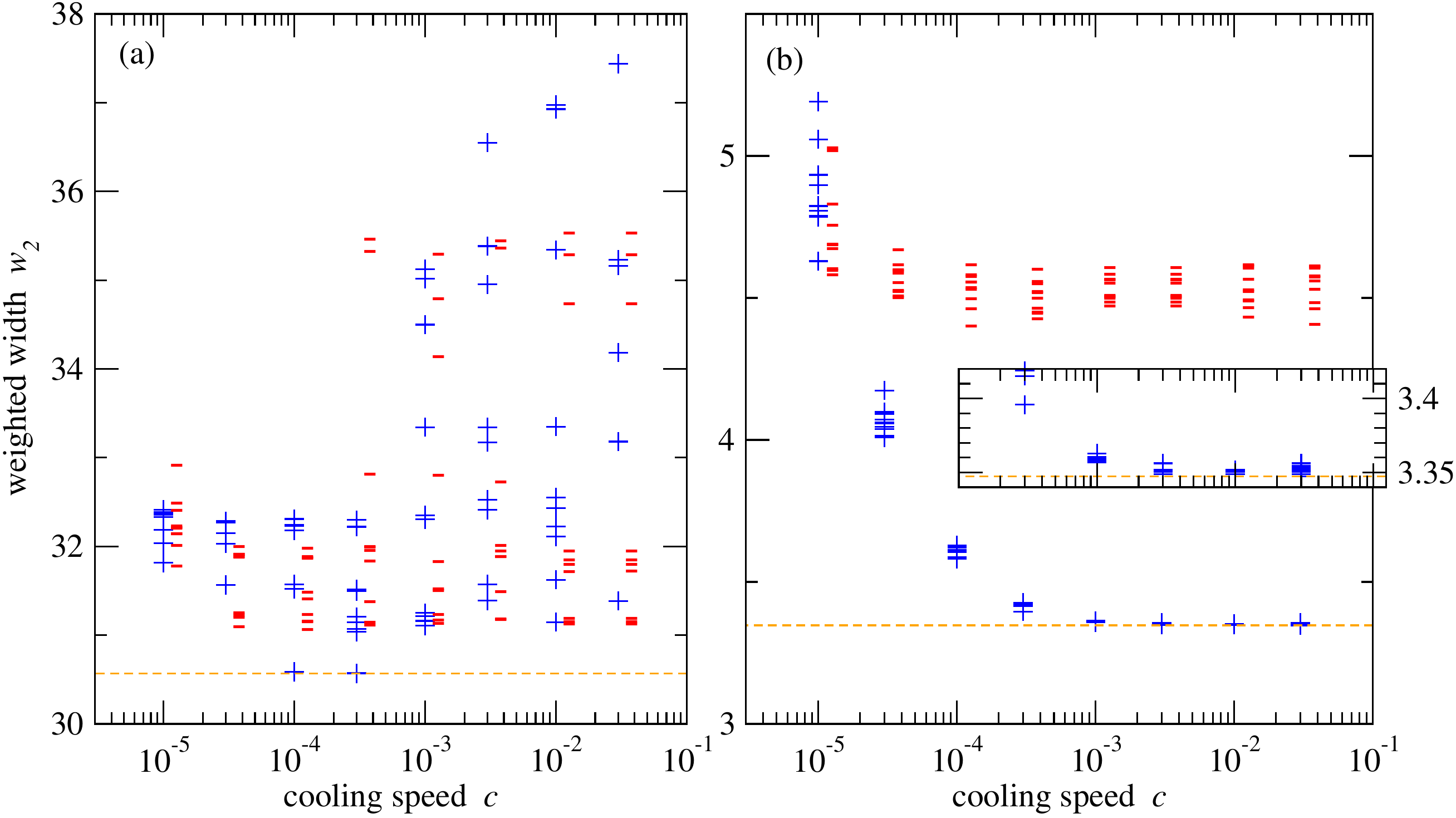}}
\caption{\label{fig:scatter}
Results of simulated annealing on networks {\tt college football} in panel (a), and {\tt network science} in panel (b). Each plotted data point is the lowest value of the weighted width $w_2$ encountered during a run of $t=10^6$ sweeps. For each choice of cooling speed $c$, there are 10 independent runs under cost function $w_2(\pi)$ (symbol $+$) and another 10 runs using cost function $m$ (symbol $-$) in the acceptance step. For both cost functions, the cooling speed takes values $c \in \{ 10^{-5}, 3 \times 10^{-4}, 10^{-4}, \dots, 3 \times 10^{-2} \}$. Symbols $-$ are shifted to the right for better visibility. Horizontal dashed lines indicate the minimum of $w_2$ over all runs, being $30.5677$ for {\tt college football} and $3.3472$ for {\tt network science}. Cost functions are defined in Section \ref{subsec:costfunc}.
}
\end{figure}

As a first step in exploring elimination orders by simulated annealing, we compare the effects of the choice of cost function, see also Section \ref{subsec:costfunc}. Figure \ref{fig:scatter} is a scatter plot of minimal $w_2$ values obtained in simulated annealing runs. Better performance is obtained with optimization under cost function $w_2$ as compared to cost function $m$. The following optimization runs on all networks in the data set are thus performed with cost function $w_2$ only.

%\subsection{Tree-width estimates} %%%%%%%%%%%%%%%%%%%%%%%%%%%%%%%%%%%%

\begin{table}
  \caption{\label{tab:mainres}
  Networks considered and results obtained. Next to network name and reference, columns $w_2$ and $w_\infty$ give the lowest values found by . simulated annealing. For each network, minima of $w_2$ and $w_\infty$ are taken from $10^6$ sweeps of 80 independent runs, 10 runs for each of cooling speed value $c \in \{ 10^{-5}, 3 \times 10^{-4}, 10^{-4}, \dots, 3 \times 10^{-2} \}$. The rightmost column is the exact number $b$ of edges in a maximum cut (edge-maximal bipartite subgraph) on the network.
Networks have $|V|$ nodes and $|E|$ edges.
  }
  \centerline{\begin{tabular}{|l|r|r|r|r|r|r|} \hline
  Network         & ref.               & $w_2$  & $w_\infty$ & $|V|$ & $|E|$ & $b$ \\ \hline 
  Karate club     & \cite{Zachary:1977}& 3.0572 &   5        &  34   &   78  &  61 \\ \hline % 0.7821
  Social 3        & \cite{Milo:2004}   & 5.3208 &   7        &  32   &   80  &  61 \\ \hline % 0.7625 
  Protein 2       & \cite{Milo:2004}   & 3.7212 &   6        &  53   &  123  &  92 \\ \hline % 0.7480
  Social 1        & \cite{Milo:2004}   & 6.2729 &  10        &  67   &  142  & 111 \\ \hline % 0.7820
  Dolphins        & \cite{Lusseau:2003}& 6.4244 &  10        &  62   &  159  & 122 \\ \hline % 0.7673
  S 208           & \cite{Milo:2002}   & 3.6738 &   6        & 122   &  189  & 168 \\ \hline % 0.8889
  E. Coli, transc.& \cite{Mangan:2003} & 4.3316 &   7        &  97   &  212  & 163 \\ \hline % 0.7689
  Protein 1       & \cite{Milo:2004}   & 3.5963 &   5        &  95   &  213  & 160 \\ \hline % 0.7512
  Les Miserables  & \cite{Knuth:1993}  & 5.6124 &   9        &  77   &  254  & 169 \\ \hline % 0.6654
  College football& \cite{Girvan:2002} &30.5677 &  34        & 115   &  316  &     \\ \hline % 
  S 420           & \cite{Milo:2002}   & 4.6018 &   9        & 252   &  399  & 354 \\ \hline % 0.8872
  Political Books & \cite{Krebs:2004}  & 9.4579 &  13        & 105   &  441  & 309 \\ \hline % 0.7007
  David Copperfield& \cite{Newman:2006}&23.8955 &  28        & 112   &  425  & 330 \\ \hline % 0.7765
  S 838           & \cite{Milo:2002}   & 5.2136 &   9        & 512   &  819  & 724\\ \hline  % 0.8840
  Network Science & \cite{Newman:2006} & 3.3472 &   8        & 379   &  914  & 636\\ \hline  % 0.6958
  \end{tabular}
  }
\end{table}

Table \ref{tab:mainres} is the overview of results for all 15 networks considered. For 13 out of these, we find tree decompositions with $w_2 <10$ and $w_\infty \le 13$. For the networks {\em College football} and {\em David Copperfield}, the best tree decompositions found are wider.  

The tree decompositions obtained by optimization enable us to calculate the exact maximum cut size for 14 out of the 15 networks. For the network {\tt College football} with the tree decomposition of width 34, the computation runs out of memory on the machine currently used. The fraction of edges $b / |E|$ contained in a maximum cut is between $0.75$ and $0.80$ for most of the networks. Exceptions are the electronic circuits {\tt S 208}, {\tt S 420}, and {\tt S 838} with $b / |E| \approx 0.888$, confirming these networks to be close to bipartite \cite{Milo:2002}.

\section{Discussion}

Originating in computer science and discrete mathematics, the concept of tree decompositions helps to exploit sparseness and tree-like structure in the analysis of network systems. The purpose of this contribution is an exposition of tree decompositions from the perspective of complex systems in physics. There are three results. (i) A simple method for finding tree decompositions by simulated annealing, tested on real networks; (ii) the resulting tree decompositions, encoded as node elimination orders, see the supplement \cite{Supplement}; and (iii) an example of an exact computation (maximum cut) made feasible by the use of tree decompositions. 

Resulting from simulated annealing, the best tree decompositions we found are not necessarily optimal ones yet. In fact, the annealing method may be improved to become more suitable for the particular search space, e.g.~with an adaptive cooling schedule \cite{Henderson:2003}. Sometimes pictured as simulated annealing with ongoing restart, parallel tempering \cite{Earl:2005} may lead to better tree decompositions. Another ingredient for the method is to also compute a lower bound on tree-width \cite{Bodlaender:2011}. Then the search for a good tree decomposition can be stopped once we are sufficiently close to the lower bound.

Tree decompositions enable us to efficiently analyze results for networks with respect to processes running on them. The method for the maximum-cut size and maximum Ising energy presented here is the basis for obtaining the exact partition function and complete density of states for Ising and Potts models \cite{Klemm:2020}. For models of epidemic spreading and percolation \cite{GrassbergerMathbiosc:1983}, tree decompositions support the computation of cluster size distributions and epidemic thresholds as well \cite{PontSerra:2017,PontSerra:2020}.

\bigskip

Funding from MINECO through the Ram{\'o}n y Cajal program and through project SPASIMM, FIS2016-80067-P (AEI/FEDER, EU) is acknowledged.

\bigskip

\bibliographystyle{unsrt}
\bibliography{elorder}

\end{document}